\documentclass[%
 reprint,longbibliography,preprintnumbers,
nofootinbib,
 amsmath,amssymb,
 aps
]{revtex4-1}
\pdfoutput=1
\usepackage{graphicx}
\usepackage[utf8]{inputenc}
\usepackage{flushend}
\usepackage{dcolumn}
\usepackage{bm}
\usepackage{balance}

\usepackage[normalem]{ulem}

\usepackage[colorlinks = true,
            linkcolor = blue,
            urlcolor  = blue,
            citecolor =green,
            anchorcolor = blue]{hyperref}
\usepackage{verbatim}
\usepackage{color,ulem}
\usepackage[english]{babel}

\usepackage[utf8]{inputenc}
\input Starburst.fd
\newcommand*\initfamily{\usefont{U}{Starburst}{xl}{n}}\initfamily

\newcommand{\beq}{\begin{eqnarray}}
\newcommand{\eeq}{\end{eqnarray}}
\usepackage{amsmath}
\usepackage{tikz}
\usetikzlibrary{decorations.pathmorphing}
\usetikzlibrary{shapes.misc}
\tikzset{cross/.style={cross out, draw=black, minimum size=8*(#1-\pgflinewidth), inner sep=0pt, outer sep=0pt},
cross/.default={1pt}}
\usetikzlibrary{patterns,math}
\begin{document}

\title{\Large Reply to Chen and Ni on ``Explicit analytical solution for random close packing in $d=2$ and $d=3$''}

\author{\textbf{Alessio Zaccone}$^{1,2}$}%
 \email{alessio.zaccone@unimi.it}
 
 \vspace{1cm}
 
\affiliation{$^{1}$Department of Physics ``A. Pontremoli'', University of Milan, via Celoria 16,
20133 Milan, Italy.}
\affiliation{$^{2}$Cavendish Laboratory, University of Cambridge, JJ Thomson
Avenue, CB30HE Cambridge, U.K.}


\maketitle
In a quick response to our recent work \cite{Zaccone2022} on an analytical derivation of the random close packing (RCP) density in $d=2$ and $d=3$ based on statistical arguments due to liquid-theory combined with marginal stability \cite{Scossa}, Chen and Ni \cite{Chen_Comment} present some considerations, which can be summarized as follows:
(i) CN criticize the use of liquid-state theories (Percus-Yevick and/or Carnahan Starling expressions) for the radial distribution function (rdf) as they have been used in our derivation in a regime where they may lead to big errors; (ii) they claim that our derivation based on using a Dirac-delta contribution combined with the rdf from liquid theories is unjustified.

Regarding point (i): CN's Comment does not bring any new information.
In our paper, it was already very clearly stated that (quoting directly from \cite{Zaccone2022}, bottom of 1st column on page 2): 
``To deal with the strong statistical correlations among particles in the dense hard sphere system, we employ suitably modified liquid state theory for the radial distribution function (RDF). It is known that liquid theories of the RDF are unable to predict the divergence of pressure at RCP, and also cannot predict the formation of permanent nearest-neighbor contacts at RCP. However, they still provide a useful analytical starting point to account for the statistical increase of crowding around a test particle, as $\phi$ increases [20].''

The fact that this ``extrapolation'' of statistical liquid theory was already openly declared in our paper \cite{Zaccone2022}, and adequately emphasized therein, makes CN's point (i) unnecessary.

Furthermore, CN's Comment contains an incorrect statement: they claim that PY or CS theories for $g(\sigma)$ were never used at such high packing fractions $\phi$ as those of jamming or FCC crystals. This is not true, of course: PY and CS theories for $g(\sigma)$ have often been used and plotted over the entire range of $\phi$, from $\phi=0$ even up to $\phi=1$, cfr. Fig. 1 in an authoritative paper by Stratt and co-workers \cite{Stratt}.
Also, their statement ``the PY or
CS equation of state has only one input $\phi$ and it does not distinguish between different structures at the same $\phi$.'' this is also no new information, as it was already stated in our paper \cite{Zaccone2022} (in the conclusion) that: ``The derivation is therefore “order agnostic” in the sense that it does not specify the structural ordering of the particles but merely their statistical excluded volume correlations.''

Regarding CN's claim (ii): they state that the rdf of PY or CS theories does not present any Dirac-delta contribution at contact. This is obvious to anyone who works in the field and, once again, already clearly stated in our paper \cite{Zaccone2022}: ``It is known that liquid theories of the RDF [...] cannot predict the formation of permanent nearest-neighbor contacts at RCP.''. 
Hence, no new information in \cite{Chen_Comment} in this respect, either.

In our paper, we used the standard form of rdf for jammed packings with a Dirac-delta at contact, cfr. Eq. (17) in \cite{Torquato_2018} and we employed $g(\sigma)$ from PY or CS theories, as stated clearly in our paper, as ``a useful analytical starting point to account for the statistical increase of crowding around a test particle, as $\phi$ increases.'', quoting directly from \cite{Zaccone2022}.
This is a physically meaningful way of analytically constructing the rdf near jamming: the rational is that, as long as the system is below jamming, of course, as is well-known, there are no permanent contacts and $z=0$, indeed. In this regime, $g(\sigma)$ only tells the number of \emph{transient} particle contacts, which clearly increases upon increasing $\phi$. As soon as the system jams, however, $z$ jumps from zero to $z=6$ at the onset of shear rigidity. The achieved mechanical stability at jamming/RCP (due to affine elasticity overcoming negative/softening nonaffine contributions, as explained in \cite{Scossa}) thus causes the transient neighbours to become permanent or long lived, hence the Dirac-delta contribution at RCP.

Further justification for this approach has been given in \cite{Zaccone2022} \emph{a posteriori}, by showing the capability of the derivation to recover sensible values of packing fractions for the jammed RCP states, and it is unclear what further justification would be needed on top of this.
This way of justifying one's approach is perfectly legitimate in physics and has been used many times as a crucial step in important physical theories in the past.

Therefore, also in this case, the answer to CN's query was already contained in \cite{Zaccone2022}.

In conclusion, all of CN's critiques \cite{Chen_Comment} were already preemptively answered in our paper \cite{Zaccone2022}.
Some incorrect statements in CN's Comment have been pointed out above.
The Comment \cite{Chen_Comment} does not contain any new information, and raises questions for which very clear answers can already be found in our original paper \cite{Zaccone2022}.

\bibliographystyle{apsrev4-1}

\bibliography{refs}

\end{document}